\documentclass[%
 reprint,
 amsmath,amssymb,
 aps,
prb,
]{revtex4-1}

\usepackage{graphicx}
\usepackage{dcolumn}
\usepackage{bm}
\usepackage{mathptmx}
\usepackage{amsmath}
\usepackage{braket}
\usepackage{color}
\usepackage{pdfpages}
\usepackage{pgffor}
\usepackage{etoolbox} 

\makeatletter
\patchcmd{\@outputpage@head}{\@ifx{\LS@rot\@undefined}{}{\LS@rot}}{}{}{}
\makeatother

\begin{document}


\title{Theoretical Characterization of Structural Disorder in the Tetramer Model Structure of Eumelanin}

\author{Oleg Sapunkov}
\affiliation{%
 Department of Mechanical Engineering, Carnegie Mellon University.\\
}

\author{Abhishek Khetan}%
\affiliation{%
 Department of Mechanical Engineering, Carnegie Mellon University.\\
}
\author{Vikram Pande}%
\affiliation{%
 Department of Mechanical Engineering, Carnegie Mellon University.\\
}
\author{Venkatasubramanian Viswanathan}%
 \email{venkvis@cmu.edu}
\affiliation{%
 Department of Mechanical Engineering, Carnegie Mellon University.\\
}%


%

\date{\today}

\begin{abstract}

Eumelanin is regarded to be an attractive candidate material for biomedical applications. Despite many theoretical studies exploring the structure of eumelanin, an exact mapping of the energetic landscape of the very large phase space of eumelanin is still elusive.  In this work, we implement a piecewise Ising Model to predict formation enthalpies of Eumelanin single and double tetramers, and demonstrate its superior predictive and generalizable capabilities. We believe this model will prove very useful in theoretically characterizing the many unique properties attributed to its disorder. The modular nature of the predictive Ising model built up in this work is well-suited for analysis and characterization of a larger phase space of eumelanin polymers, including hexamers and octomers, as well as larger stacked structures, such as potential triple and quadruple eumelanin tetramers. Absorbance data can be incorporated with population-wide predictions of polymer abundance to produce weighted-average predictions of broadband absorbance of bulk eumelanin.

\end{abstract}

\pacs{Valid PACS appear here}
\maketitle


\section{Introduction}

Eumelanin is a subgroup of melanin pigments found in living organisms that plays an important role in skin coloration and UV protection.\cite{friedmann1987ultraviolet} The precise chemical structure of eumelanin is still not completely known because it is highly cross-linked and insoluble in available solvents.\cite{ito2013high} Among organic polymers, eumelanin occupies a unique position because of: (i) its widespread occurrence in nature, from people and mammals to fish, birds and molluscs;\cite{edwards2014pigments} (ii) the variety of biological roles, from photoprotection to scavenging of reactive oxygen species\cite{nofsinger2002aggregation, bustamante1993role} and metal chelation;\cite{im2017metal} and (iii) distinct physical and chemical properties, including broadband photoabsorption throughout the visible range,\cite{tran2006chemical} water-dependent ionic-electronic semiconductor-like behavior,\cite{meredith2006physical} stable free radical character and efficient nonradiative energy dissipation,\cite{ju2011bioinspired} making eumelanin an attractive candidate for biomedical and technological applications. Despite growing interest in eumelanin-type functional materials and systems, the exact structural underpinnings due to the highly insoluble and heterogeneous character of these polymers, has proved challenging.\cite{d2009chemical}

In parallel, interest in miniaturized medical implants and edible biometric sensors has led to a need to develop novel biodegradeable batteries, based on eumelanin extracted from the common cuttlefish ($Sepia~officinalis$)\cite{kim2016evidence, kim2013biologically}.  Electrochemical characterization indicates the possibility of electrochemical intercalation of up to two sodium ions per eumelanin unit.\cite{kim2016evidence}  The theoretical analysis in the work of Kim et al.\cite{kim2016evidence} utilizes a stacked tetramer model to rationalize these findings, originally proposed by Kaxiras \textit{et al}.\cite{kaxiras2006structural}  Recently, another study has explored the geometric complexity possible within the Kaxiras model.\cite{di2017natural, vahidzadeh2018melanin, crescenzi2017kaxiras} While these studies represent important strides, a complete energetic landscape of double tetramer is still elusive given the large phase space.

Systematic exploration of large phase spaces for crystalline materials is enabled through the Ising Model (or cluster expansion).\cite{ceder1993derivation, de1992asymmetric, tepesch1995model} The general methodology involves calculating the system energy for a subset of structures and training a model that can be used to then subsequently predict the rest of the phase space with very high accuracy.  In this work, we develop an Ising model to describe the interactions in-plane and out-of-plane for a double tetramer.  Utilizing 108 density functional theory calculations, we train an Ising model that predicts on an test set with an accuracy of $0.15$ eV.  The model is generalizable and allows for an accurate mapping of the energetic landscape of a double tetramer.  We believe this analysis can form the basis for further characterization of broadband absorbance, electrochemical ion interacalation, etc.



\section{Methods}

\subsection{DFT Simulations and Structure}

In this work, the primary molecular structure considered for this study is a double eumelanin tetramer, building on the work of Kaxiras \textit{et al}.\cite{kaxiras2006structural} While other possible structures such as hexamer\cite{randhawa2009evidence, stark2005effect} and octomer\cite{slominski2004melanin} are important, we have chosen to utilize the double tetramer structural model to demonstrate our methodology.  As will be discussed later, this model can be extended to other possible structures of eumelanin.

\begin{figure*}[!htb]
 \centering
 \includegraphics[width=0.75\textwidth]{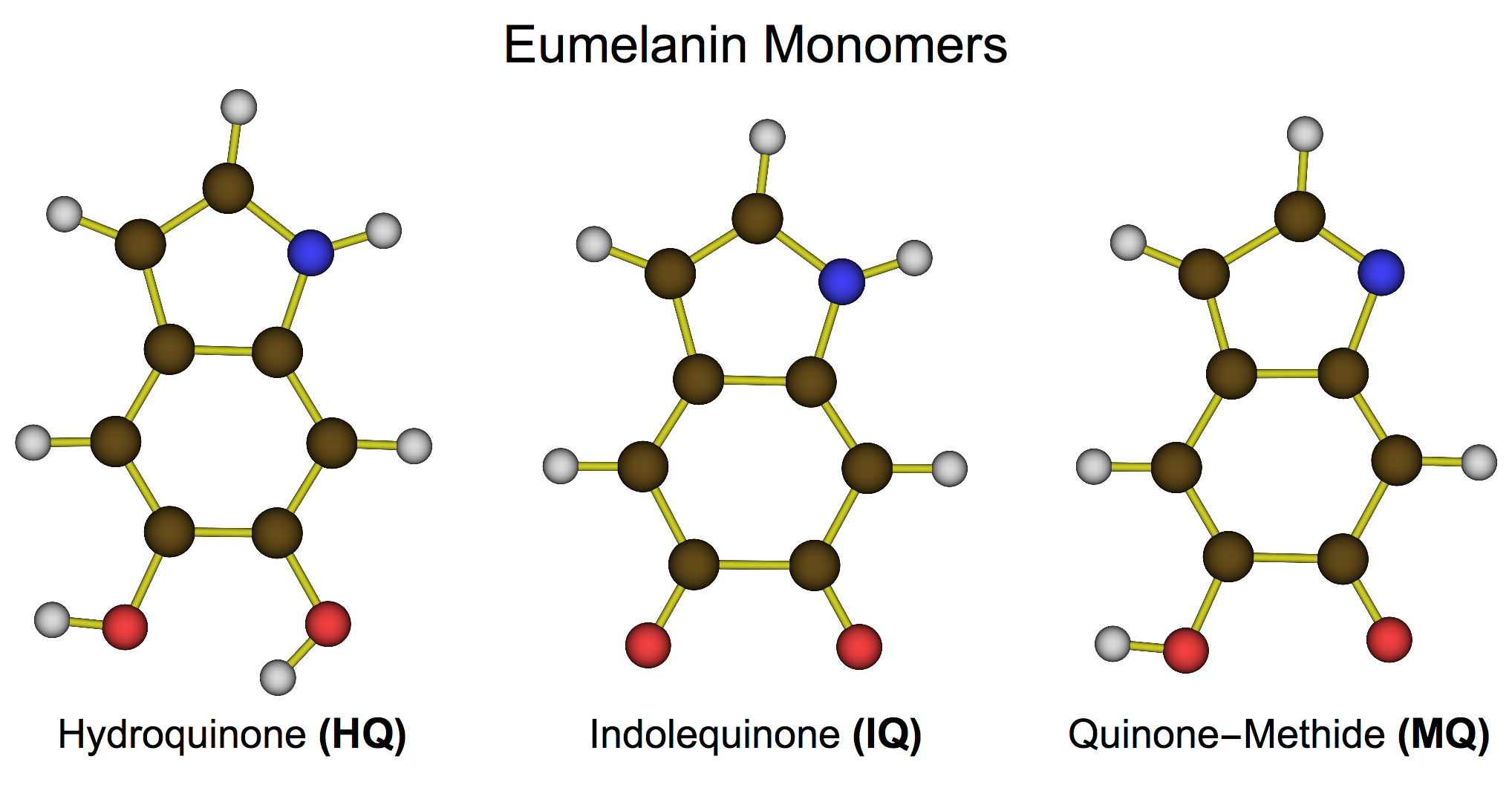}
\caption{Structures of the 3 primary eumelanin monomers used to form single and double tetramers investigated in the present study.}
\label{fgr:EumelaninM}
\end{figure*}

The specific structures simulated for this study includes 3 types of eumelanin monomers: hydroquinone (HQ), indolequinone (IQ), and quinone-methide (MQ).\cite{meng2008theoretical} The fourth and final monomer type described by Kaxiras, quinone-imine (NQ), was not used in this study, since it is an isomer of quinone-methide (MQ). In the context of planar tetramer assembly and ion intercalation, its structural difference is negligible, and thus it is expected to produce structures with similar formation enthalpies. The monomer structures are represented in Fig. \ref{fgr:EumelaninM}.

\begin{figure*}[!htb]
 \centering
 \includegraphics[width=0.75\textwidth]{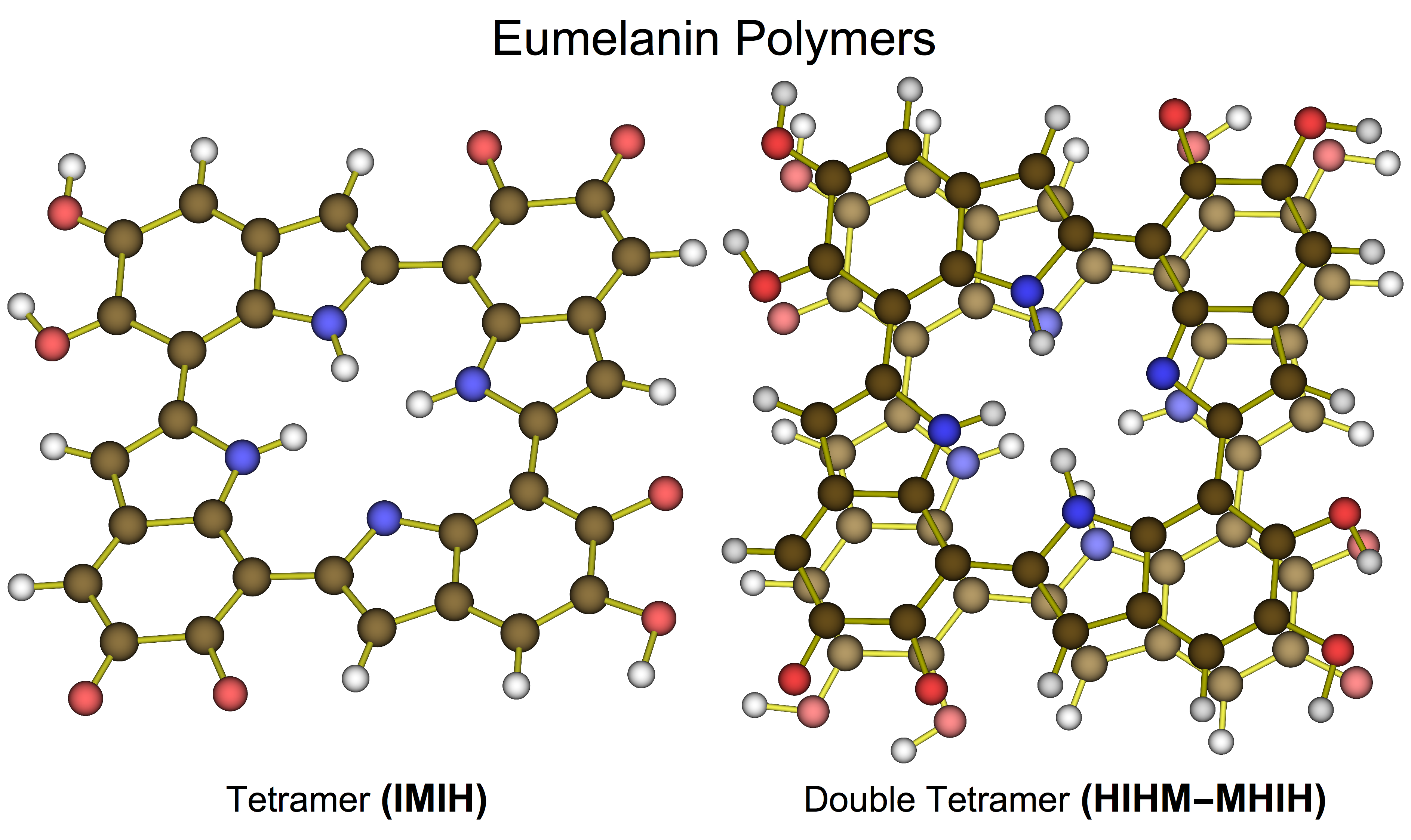}
\caption{Examples of a Eumelanin tetramer and Eumelanin double tetramer, showing the planar arrangement of monomers within tetramers and the parallel sheet stacking of tetramers.}
\label{fgr:EumelaninPM}
\end{figure*}

We introduce the following naming scheme to describe the geometric arrangement of simulated structures. The double eumelanin tetramers are planar in the X-Y plane, stacked one above the other along the Z axis. This is one of the stable stacked structures when vdW interactions are taken into account.\cite{kim2016evidence} Looking down the Z axis onto the X-Y plane, the monomers of the lower tetramer are named first, starting with the monomer in Quadrant 2 and proceeding clockwise. Following the 4 monomers of the lower tetramer, the 4 monomers of the upper tetramer are named, in the same order. All monomers are designated by the first letter of their acronym (H for hydroquinone, I for indolequinone, M for quinone-methide). Thus, for instance, in an example tetramer HHHM-HHHM, the 2 sets of 3 hydroquinones, in Quadrants 2, 1, and 4, are situated one above the other, and the 2 sets of single quinone-methides, in Quadrant 3, are likewise situated one above the other. Fig. \ref{fgr:EumelaninPM} illustrates the structures of an example planar tetramer and an example double tetramer.

In addition to simulations of the double eumelanin tetramer, single eumelanin planar tetramers were also simulated, to compute formation enthalpies of the double tetramers from single tetramers. As will be discussed later, the purpose of these calculations was to ensure the out-of-plane interactions between the simulated pairs of planar double tetramers were properly quantified and accounted for.

Finally, single eumelanin monomers were also simulated, to compute formation enthalpies of both the single and double eumelanin tetramers. These single monomers were simulated in their stable standalone configuration, with an additional 2 hydrogen atoms, which are lost in the polymerization process into tetramers, retained on the monomers.

Self-consistent DFT calculations were performed using the Projector Augmented Wave Method as implemented in GPAW,\cite{enkovaara2010electronic} with the Bayesian Error Estimation Functional\cite{mortensen2005bayesian} with van-der-Waals correlation\cite{wellendorff2012density} (BEEF-vdW) exchange-correlation functional. All molecules were simulated in standalone bulk cells measuring 20 x 20 x 17 {\AA}, with all periodic boundary conditions disabled. Calculations were run with a real-space grid of 0.18 {\AA} The Conjugate Gradient eigensolver was used to facilitate convergence of the simulations. Fermi-Dirac occupation smearing of 0.01 eV was used to expedite simulation convergence while providing accurate results. All structures were geometrically relaxed until the net force in the simulated molecule was decreased to 0.05 eV.

\subsection{Formation Enthalpies from Bulk Simulations}

Several sets of formation enthalpies were calculated for the assemblage of simulated structures, including relevant monomers, planar tetramers, and double tetramers.

First, formation enthalpies of only the double tetramers (T-T) were calculated, using DFT-calculated internal energies of the double tetramers and DFT-calculated reference energies for the constituent elemental species: carbon, hydrogen, nitrogen, and oxygen.
\begin{equation}
{\Delta}H_{T-T}^{a} = U_{T-T}^{DFT} - \sum_{C} E_{C}^{Ref} - \sum_{H} E_{H}^{Ref} - \sum_{N} E_{N}^{Ref} - \sum_{O} E_{O}^{Ref} + \Delta pV
\end{equation}
The pressure-volume work term, $\mathrm{\Delta pV}$, can be disregarded, as it is typically about 5 orders of magnitude smaller than internal energy contributions in formation enthalpy calculations.\cite{aydinol1997ab,obrovac2014alloy}

Elemental reference energy for hydrogen was calculated using simply the DFT-calculated internal energy of hydrogen gas, H$_{2}$. Elemental reference energy for carbon was calculated using bulk graphite.\cite{pande2018robust} The reference energies for oxygen and nitrogen, however, required additional correction, since both are well-known to be poorly described within DFT.\cite{jacob2006mechanism, rod1999nitrogen} Oxygen reference energy was calculated using the standard water-reference scheme,\cite{baran2014analysis} using DFT-computed internal energies of water and hydrogen gas, as well as the experimental formation enthalpy of water:
\begin{equation}
E_{O}^{Ref} = U_{H_{2}O}^{DFT} - U_{H_{2}}^{DFT} - {\Delta}H_{H_{2}O}^{Exp}
\end{equation}
Nitrogen reference energy was calculated using a similar, ammonia-reference scheme,\cite{wang2011controllable} using DFT-computed internal energies of ammonia and hydrogen gas, as well as the experimental formation enthalpy of ammonia:
\begin{equation}
E_{N}^{Ref} = U_{NH_{3}}^{DFT} - {3/2} U_{H_{2}}^{DFT} - {\Delta}H_{NH_{3}}^{Exp}
\end{equation}

In addition to calculating formation enthalpies of the double tetramers from the elemental basis, formation enthalpies of the single planar tetramers (T) and the single monomers (M) were also computed, using the same elemental basis.
\begin{equation}
{\Delta}H_{T}^{a} = U_{T}^{DFT} - \sum_{C} E_{C}^{Ref} - \sum_{H} E_{H}^{Ref} - \sum_{N} E_{N}^{Ref} - \sum_{O} E_{O}^{Ref}
\end{equation}
\begin{equation}
{\Delta}H_{M}^{a} = U_{M}^{DFT} - \sum_{C} E_{C}^{Ref} - \sum_{H} E_{H}^{Ref} - \sum_{N} E_{N}^{Ref} - \sum_{O} E_{O}^{Ref}
\end{equation}

\subsection{Ising Model Coefficients}

In order to map out the energetic interactions between the eumelanin monomers, both within the planar tetramers and across the planes of the double tetramers, we utilize a modified Ising Model for the lattice Hamiltonian. As implemented, the model consists of a lattice of $N$ sites $i$, whose filling is described by occupation terms, $\sigma_{i}$. All terms within the Ising Model are calculated using formation enthalpies of the monomers and polymers, referenced to their constituent atomic species, as described above. The occupation energies, $h$, in this implementation of the Ising Model, represent the formation enthalpies of the monomers directly. As there are 3 types of monomers used in the explored configuration space of eumelanin double tetramers, 3 types of occupation terms were used: $h_{H}$, $h_{I}$ and $h_{M}$, each term corresponding to the formation enthalpy of its corresponding monomer. In addition to the occupation terms $h$, energy contributions to the full polymer due to monomer-monomer interactions were captured by the interaction terms $j$. Several types of interaction terms are identified, separated into in-plane interactions between monomers situated in the same planar tetramer, $j_{i}$, and out-of-plane interactions between monomers situated in opposing tetramers, $j_{o}$. Within each category, interactions between monomers occupying nearby quadrants are accounted for, $j_{in}$ and $j_{on}$, as well as interactions between monomers occupying diametrically opposed quadrants, $j_{id}$ and $j_{od}$. Finally, interactions between monomers situated in opposing dimers but in the same geometric quadrant, one underneath another, are accounted for as $j_{ou}$. Thus, a total of 5 major types of interaction terms are used in the Ising Model: $j_{in}$, $j_{id}$, $j_{ou}$, $j_{on}$, $j_{od}$. Within each type of interaction term, there is further distinction as to the types of monomers involved, with the full set of permutations between H, I, and M accounted for: HH, HI, HM, II, IM, and MM. 

Two schemes were used to calculate the above occupation and coupling terms for the Ising Model. In the first scheme, the formation enthalpy data of eumelanin double tetramers, calculated through DFT simulations, was used to derive the full set of corresponding Ising Model coefficients in a single step, though a least-squares regression fit using Wolfram Mathematica. Each double tetramer formation enthalpy was described as:
\begin{multline}
{\Delta}H_{T-T}^{a} = \sum_{\braket{i}} h_{i}\sigma_{i} + \sum_{\braket{ik}} j_{in_{i,k}}\sigma_{i}\sigma_{k} + \sum_{\braket{ik}} j_{id_{i,k}}\sigma_{i}\sigma_{k} + \\ + \sum_{\braket{ik}} j_{ou_{i,k}}\sigma_{i}\sigma_{k} + \sum_{\braket{ik}} j_{on_{i,k}}\sigma_{i}\sigma_{k} + \sum_{\braket{ik}} j_{od_{i,k}}\sigma_{i}\sigma_{k}
\end{multline}

The second scheme broke down the calculation of Ising Model coefficients into three distinct steps, using the formation enthalpies of the monomers, single tetramers, and double tetramers for each step, respectively. First, the formation enthalpies of the monomers were used to calculate the occupation terms $h$ directly. A correction had to be implemented for hydrogen, since each monomer lost two hydrogen atoms when it was polymerized into a tetramer:
\begin{equation}
{\Delta}H_{M_{i}}^{a} - 2 E_{H}^{Ref} = h_{i}\sigma_{i}
\end{equation}

Next, the formation enthalpies of the single planar tetramers were used to calculate the in-plane coupling coefficients $j_{i}$. The formation enthalpy of a single tetramer contains contributions both from the presence of individual monomers and their in-plane interactions:
\begin{equation}
{\Delta}H_{T}^{a} = \sum_{\braket{i}} h_{i}\sigma_{i} + \sum_{\braket{ik}} j_{in_{i,k}}\sigma_{i}\sigma_{k} + \sum_{\braket{ik}} j_{id_{i,k}}\sigma_{i}\sigma_{k}
\end{equation}
To isolate the energetic contributions of in-plane interactions, we subtracted the occupation-energy contributions of the constituent monomers, which were quantified in the preceding step, from the formation enthalpy of the tetramer. In effect, the formation enthalpy of a tetramer from the monomer basis was calculated. The resultant energy was used to calculate the in-plane coupling coefficients, though a least-squares regression fit using Wolfram Mathematica:
\begin{equation}
{\Delta}H_{T}^{a} - \sum_{\braket{i}} h_{i}\sigma_{i} = {\Delta}H_{T}^{M}
\end{equation}
\begin{equation}
{\Delta}H_{T}^{M} = \sum_{\braket{ik}} j_{in_{i,k}}\sigma_{i}\sigma_{k} + \sum_{\braket{ik}} j_{id_{i,k}}\sigma_{i}\sigma_{k}
\end{equation}

Finally, the formation enthalpies of the double tetramers were used to calculate the out-of-plane coupling coefficients $j_{o}$. As discussed in the one-step scheme, the formation enthalpy of the double tetramer contains information about all types occupation and coupling terms, so we subtracted the occupation-energy contributions, calculated using the monomers, and the in-plane interaction-energy contributions, calculated using the single tetramers. Effectively, the remaining energy represented the formation enthalpy of a double tetramer from its constituent single tetramers. This formation enthalpy was used to fit the out-of-plane interaction coefficients, though a least-squares regression fit using Wolfram Mathematica:
\begin{equation}
{\Delta}H_{T-T}^{a} - ( \sum_{\braket{i}} h_{i}\sigma_{i} + \sum_{\braket{ik}} j_{in_{i,k}}\sigma_{i}\sigma_{k} + \sum_{\braket{ik}} j_{id_{i,k}}\sigma_{i}\sigma_{k} ) = {\Delta}H_{T-T}^{T}
\end{equation}
\begin{equation}
{\Delta}H_{T-T}^{T} = \sum_{\braket{ik}} j_{ou_{i,k}}\sigma_{i}\sigma_{k} + \sum_{\braket{ik}} j_{on_{i,k}}\sigma_{i}\sigma_{k} + \sum_{\braket{ik}} j_{od_{i,k}}\sigma_{i}\sigma_{k}
\end{equation}

\subsection{Double Tetramer Phase Space}

Following the derivation of the interaction coefficients, these coefficients were applied to predict the formation enthalpies of the entire phase space of relevant eumelanin double tetramers. If every possible permutation of HQ, IQ, and MQ within a double tetramer is considered, there exist a total of 6561 distinctly named double tetramers. However, many of these tetramers are simply rotated or flipped equivalents of other tetramers within the same phase space. To account for this phase space degeneracy, the full set of Ising Model coupling coefficients for each of the 6561 configurations was calculated analytically, and families of configurations which matched the set of coupling coefficients had all but one member removed from the phase space. This left a total of 1032 distinct, non-degenerate double tetramer configurations in the phase space.

To provide training data for the Ising Model, a total of 85 double tetramers composed of HQ, IQ, and MQ monomers were simulated in DFT using the BEEF-vdW exchange-correlation functional. In addition, the 3 monomers themselves were simulated in DFT, as well as 20 intermediate single planar tetramers.

This study yielded two sets of interaction coefficients, as described above: one derived from the formation enthalpies of the double tetramers calculated directly from the atomic basis in one step, the other derived from the formation enthalpies of the double tetramers, single tetramers, and monomers, all calculated from the atomic basis, in three steps. The use of the BEEF-vdW exchange correlation functional generates a non-self consistent ensemble of energies, which was used to train an ensemble of Ising models.  It has been shown that the ensemble of energies reliably reproduces trends in energies at the GGA-level.\cite{houchins2017quantifying, vinogradova2018quantifying, krishnamurthy2016universality, krishnamurthy2018maximal}

To most adequately represent the distribution of predicted formation enthalpies of the melanin double tetramers, it was decided to use a histogram plot of the calculated formation enthalpies predicted for every double tetramer configuration. To produce these plots, the double tetramers were sorted in order of increasing average formation enthalpy, as calculated across the full ensemble of 2000 values predicted by the BEEF-vdW functional. For every double tetramer, the standard deviation of formation enthalpy was calculated across all 2000 predicted BEEF-vdW values, to indicate uncertainty in the calculation. This standard deviation was represented as a band on the histogram of formation enthalpies, centered on each individual column of the histogram. The band provided for each column represents the standard deviation of formation enthalpies relevant only to the double tetramer structures present in that column. Since the standard deviation slightly differed for individual double tetramers, the resultant bands are slightly but negligibly different for each column.

\section{Results $\&$ Discussion}

Using the 1-step calculation for formation enthalpy of the double tetramers, it was found that the average formation enthalpy across all simulated double tetramers, calculated from the atomic reference basis, was $-14.7$ eV, with a standard deviation of $3.1$ eV. Double tetramers containing a higher proportion of HQ (hydroquinone) were found to be the most stable, down to $-19.3$ eV, while double tetramers containing higher proportions of MQ (quinone-methide) exhibited the least negative formation enthalpies, up to $-5.22$ eV. The standard deviation of calculated formation enthalpy for each individual double tetramer, throughout the phase space of the BEEF-vdW ensemble was calculated as well. The average of these standard deviations for the set of simulated structures was $2.9$ eV, with an overall standard deviation of just $0.1$ eV.

The average formation enthalpy of an individual planar tetramer, calculated from the atomic basis, was $-5.09$ eV, with a standard deviation of $1.92$ eV. Thus, it was expected the formation enthalpy of the double tetramer from single planar tetramers would be of comparable order-of-magnitude to the formation enthalpy of the said tetramers. As calculated, the average formation enthalpy of a double tetramer from its constituent single tetramers was $-1.88$ eV, with a standard deviation of $0.58$ eV. From this observation, it was apparent that the out-of-plane coupling coefficients derived through the Ising Model should be lower than the in-plane coupling coefficients.

Finally, formation enthalpies of the individual monomers were calculated from the atomic basis. The average magnitude was found to be $-1.46$ eV, with a standard deviation of $1.11$ eV. Since each corresponding $h$ term, derived from the formation enthalpies of the monomers, had to account for the loss of 2 hydrogen atoms, the average magnitude of the occupation energies used in the 3-step Ising Model was $6.49$ eV. 

In both least-squares fits of the coupling coefficients used with the Ising Model, as fit to the training data, the reported $r^{2}$ was very high, in excess of $0.99$, for all ensemble-specific sets of coefficients. However, the values derived for the coupling coefficients differed strongly. Tab. \ref{eumelacoefficients} outlines the results summarized across each type of coupling coefficient, while Table \ref{eumelacoefficientsCOMPLETE} provides a more complete database of coefficient values, with statistics available for each individual coefficient present in the system.

\begin{table}[!htb]
\begin{tabular}{|c|c|c|c|c|}
\hline
Coeff & Grand $\mu$ (1S) & Grand $\sigma$ (1S) & Grand $\mu$ (3S) & Grand $\sigma$ (3S)\\
\hline
h & -0.28 & 0.08 & 6.49 & 1.11 \\
\hline         
jin & -0.26 & 0.23 & -4.03 & 0.24 \\
\hline         
jid & -0.46 & 0.30 & -7.40 & 0.22 \\
\hline         
jou & -0.46 & 0.37 & -0.14 & 0.17 \\
\hline         
jon & -0.20 & 0.23 & -0.05 & 0.11 \\
\hline         
jod & -0.44 & 0.04 & -0.12 & 0.06 \\
\hline
\end{tabular}
\caption{Statistics on calculated Ising Model coefficient values, in eV, summarized across each general type of coefficient. Mean of means and standard deviation of means data provided for coefficients calculated using the 1-Step and the 3-Step calculation schemes.}
\label{eumelacoefficients}
\end{table}

\begin{table}[!htb]
\begin{tabular}{|c|c|c|}
\hline
Coeff. & $\mu$ (3S) & $\sigma$ (3S) \\ 
\hline 
hH & 5.23 & 1.21 \\
\hline 
hI & 6.92 & 1.25 \\
\hline 
hM & 7.33 & 1.29 \\
\hline 
jinHH & -3.82 & 0.56 \\
\hline 
jinHI & -3.89 & 0.56 \\
\hline 
jinHM & -4.32 & 0.57 \\
\hline 
jinII & -3.88 & 0.56 \\
\hline 
jinIM & -4.37 & 0.57 \\
\hline 
jinMM & -3.92 & 0.56 \\
\hline 
jidHH & -7.16 & 1.03 \\
\hline 
jidHI & -7.23 & 1.03 \\
\hline 
jidHM & -7.55 & 1.04 \\
\hline 
jidII & -7.24 & 1.03 \\
\hline 
jidIM & -7.68 & 1.04 \\
\hline 
jidMM & -7.56 & 1.03 \\
\hline 
jouHH & -0.18 & 0.1 \\
\hline 
jouHI & -0.43 & 0.14 \\
\hline 
jouHM & -0.17 & 0.14 \\
\hline 
jouII & -0.07 & 0.07 \\
\hline 
jouIM & -0.01 & 0.11 \\
\hline 
jouMM & 0.04 & 0.09 \\
\hline 
jonHH & -0.11 & 0.06 \\
\hline 
jonHI & -0.17 & 0.05 \\
\hline 
jonHM & -0.06 & 0.06 \\
\hline 
jonII & 0.11 & 0.09 \\
\hline 
jonIM & -0.1 & 0.05 \\
\hline 
jonMM & 0.04 & 0.07 \\
\hline 
jodHH & -0.09 & 0.07 \\
\hline 
jodHI & -0.15 & 0.07 \\
\hline 
jodHM & -0.06 & 0.07 \\
\hline 
jodII & -0.08 & 0.09 \\
\hline 
jodIM & -0.11 & 0.07 \\
\hline 
jodMM & -0.23 & 0.05 \\
\hline
\end{tabular}
\caption{Statistics on calculated coefficient values, in eV, in the eumelanin Ising Model. Mean and standard deviation data provided for individual coefficients across the 2000 values exported with the BEEF-vdW ensemble, for Ising Models computed using the 3-Step calculation scheme.}
\label{eumelacoefficientsCOMPLETE}
\end{table}

It is evident that although both methods are able to match the formation enthalpies of the double tetramers well, the 1-Step calculation method provides a less physically interpretable model. In addition, this set of coupling coefficients fails to predict the formation enthalpies of single tetramers well. The predicted coefficient values using the 1-Step calculation are not generalizable as no data regarding the monomers and single-tetramers data were used in the training.  It is worth highlighting that this 1-Step analysis would constitute a black-box application of the Ising model for double tetramer model.

On the other hand, the results provided by the 3-Step calculation method are far more systematic, generalizable and physically meaningful. In the full set of fit Ising Model coefficients, it was found that the occupation coefficients and the in-plane coupling coefficients contribute the bulk of the formation enthalpy of a double tetramer, as was expected since out-of-plane interactions are dominantly vdW interactions and do not involve any chemical bonding. Of the in-plane interaction coupling coefficients, the coefficient across the diagonal of a single tetramer was found to be almost twice the magnitude of the nearest-neighbor coupling coefficient. We interpret this as indicative of the stability of the tetramer as a standalone polymer: if the nearest-neighbor coupling coefficient was much greater in magnitude than the diagonal coupling coefficient, the monomers may have preferred to form a ribbon that could stretch indefinitely, rather than forming a self-contained tetramer unit. Finally, all out-of-plane coupling coefficients were found to be much lower than the in-plane coefficients, as expected from the formation enthalpy calculations.

\begin{figure*}[!htb]
 \centering
 \includegraphics[width=0.75\textwidth]{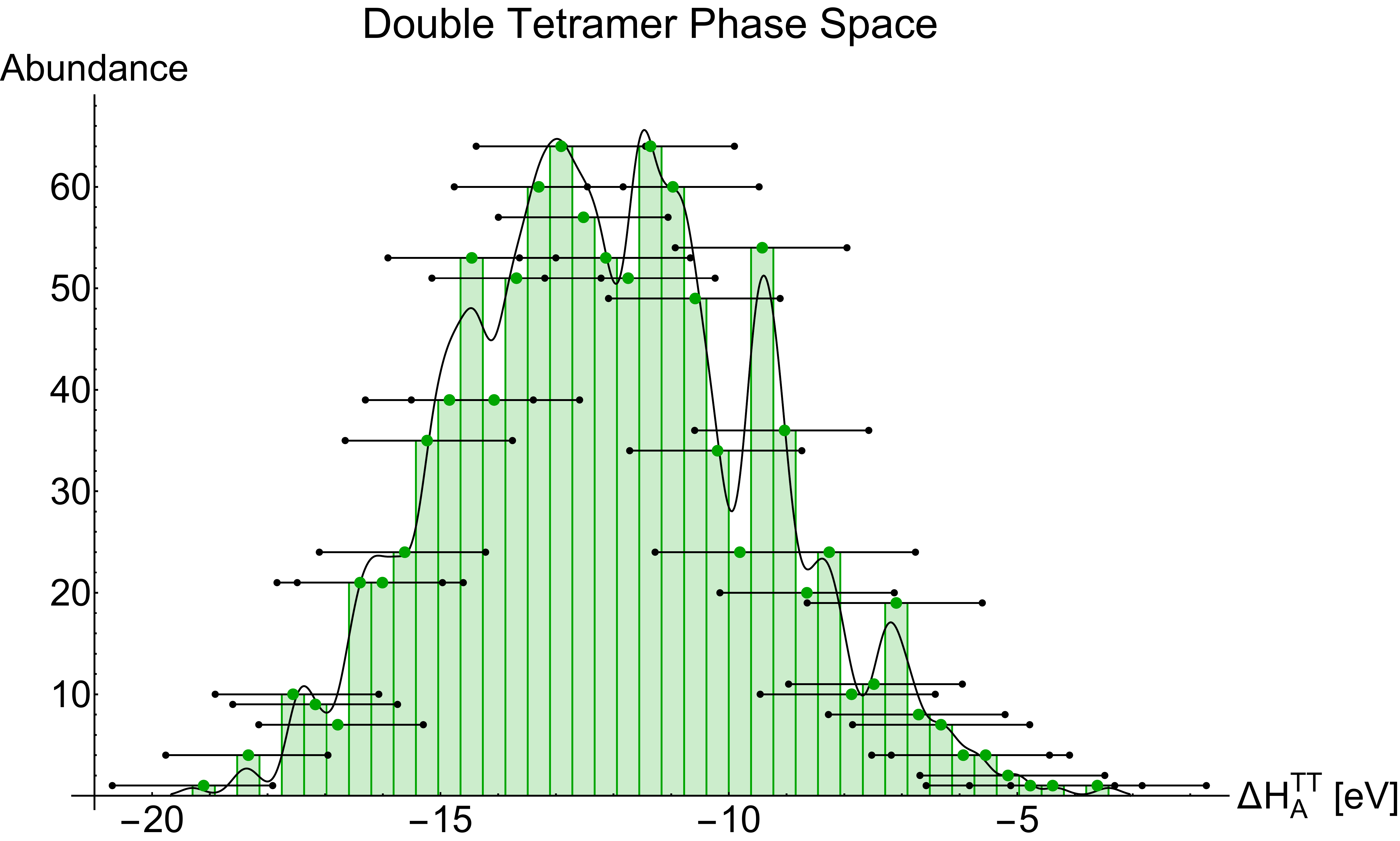}
\caption{Predicted formation enthalpies of the full double tetramer phase space using the 3-step computational method. Weak reliance on tetramer configuration can be seen, with a comparatively continuous distribution of formation enthalpies across the families of tetramers.}
\label{fgr:2StepDegen}
\end{figure*}

The effect of this disparity between the two methods can be seen in the predicted statistics regarding the expected formation enthalpies of the full double tetramer phase space. Fig. \ref{fgr:2StepDegen} shows the histogram of formation enthalpies of all double tetramers, as calculated using the 3-step computational method described above. In the plot, it is easy to see that there are some independent peaks, which correspond to double tetramer phases with a specific number of hydroquinone monomers, but the histogram overall follows a standard Bell curve. For comparison, Fig. S1 (provided with the Supporting Information document) shows the same histogram produced for the formation enthalpies calculated using the 1-step computational method. That histogram demonstrates much more significant peak formation, indicating a much stronger dependence of formation enthalpy on the presence of hydroquinone monomers, and, by extension, the out-of-plane coupling interactions, as these are magnified in the 1-step computation.

From these two formation enthalpy plots, it can be seen that the 1-Step method and the 3-Step methods generally predict the ensemble of formation enthalpies within the same range of values, both throughout the phase space of eumelanin double tetramer configuration, as well as the phase space of the BEEF ensemble. However, the characteristics of this phase space as predicted by the 1-Step and 3-Step methods differ. The 1-step method over-predicts an excessive reliance on the specific geometric configuration of the double tetramer, most likely due to its excessive emphasis on out-of-plane interactions within the phase space of tetramer geometry. The 3-step method reduces the enthalpic contribution from the out-of-plane interactions, and thereby produces a much smoother, more continuous plot. It is observed that the 3-Step method is more accurate at matching the formation enthalpies of calculated test data.

Formation enthalpy calculations showed that the single and double tetramers with the highest fraction of HQ monomers possess the lowest, most stable formation enthalpies, as compared to eumelanin structures with a higher fraction of IQ and MQ monomers. This observation was replicated throughout the entire phase space of double tetramer structures predicted by the Ising Model. This observation implies that within a mixture of eumelanin monomers polymerizing into double planar eumelanin tetramers, in the thermodynamic limit, double tetramers rich in HQ will be preferentially formed out of the available HQ population first, with only a small fraction of HQ monomers ultimately bound within IQ- and MQ-rich double tetramers. 

The coefficients calculated from the Ising Model indicate that in-plane interactions are strongest between an MQ monomer and a dissimilar monomer; both the nearest-neighbor and the diagonal next-nearest-neighbor interactions between HQ and MQ or IQ and MQ are stabilized by 0.3 - 0.5 eV compared to analogous in-plane interactions not involving MQ. This indicates that single planar tetramers formed with a small amount of MQ monomers are more stable than monomers with a mixture of HQ and IQ monomers alone. Out-of-plane coefficients calculated from the Ising Model indicate that the strongest interactions involve HQ monomers, either in combination with fellow HQ monomers or other monomers. This is expected, since the HQ-rich double tetramers were found to be the most stable structures investigated, stabilized both by the presence of HQ monomers and their out-of-plane interactions.

These observation has significant implications for the anticipated intercalation potentials of ions into these double tetramers, as one of the preferred locations for intercalation of ions is the double tetramer's inner ring. The HHHH-HHHH-type double tetramer carries 8 hydrogen atoms protruding into the inner ring, which may hinder ion intercalation, while the IQ-rich and MQ-rich double tetramers carry fewer hydrogen atoms protruding into the inner ring, and may possess strong binding centers for ion intercalation.

Recent experimental and $ab-initio$ simulation work indicates that the eumelanin tetramer and the eumelanin stacked double tetramer play an important role in metal ion binding and broadband absorption of bulk eumelanin.\cite{kim2016evidence, kaxiras2006structural} Our results arre consistent with earlier work by Kaxiras et al.\cite{kaxiras2006structural} where they show that hydroquinone-rich phases of eumelanin display are consistently more stable of the eumelanin polymers, ranging from dimers to tetramers, as compared to polymers poor in HQ. The modular nature of the predictive Ising model built up in this work is well-suited for analysis and characterization of a larger phase space of eumelanin polymers, including hexamers and octomers, as well as larger stacked structures, such as potential triple and quadruple eumelanin tetramers. Furthermore, as the model allows for phase-space-wide prediction of relative stability of specific polymers, it can be incorporated in a predictive model used to study the broadband absorbance of bulk eumelanin, by utilizing a weighted average of bulk polymer composition and calculated absorbance data.

\section{Conclusions}

In this study, we have built a generalizable, physically meaningful Ising model to describe the energetic interactions of eumelanin.  This model allows a precise mapping of the very large phase of eumelanin structures within the double tetramer model.  We highlight the importance of carrying out a step-wise training of the model coefficients using first the monomer enthalpies, followed by the single tetramer enthalpies and finally the double tetramer.  The developed model will prove to be extremely useful in rapidly exploring the absorbance, ion interacalation of eumelanin within the double tetramer model.  We also believe the developed methodology can be extended to other possible structural models for eumelanin such as the hexamer, octamer, etc.


\section{Acknowledgements}

The computational portion of this work was performed on the Arjuna computer cluster, which was funded through Carnegie Mellon College of Engineering and the departments of Mechanical Engineering, Electrical and Computer Engineering and Chemical Engineering.
The computational portion of this work was performed on Hercules computer cluster, which was funded through a Carnegie Mellon College of Engineering Equipment grant. O.B.S. acknowledges support from the DoD SMART Fellowship.
The authors thank Professor Christopher Bettinger of Carnegie Mellon University for helpful discussions on Eumelanin structure and function.

\bibliography{RSC.bib} 

\foreach \x in {1,...,4}
{%
\clearpage
\includepdf[pages={\x,{}}]{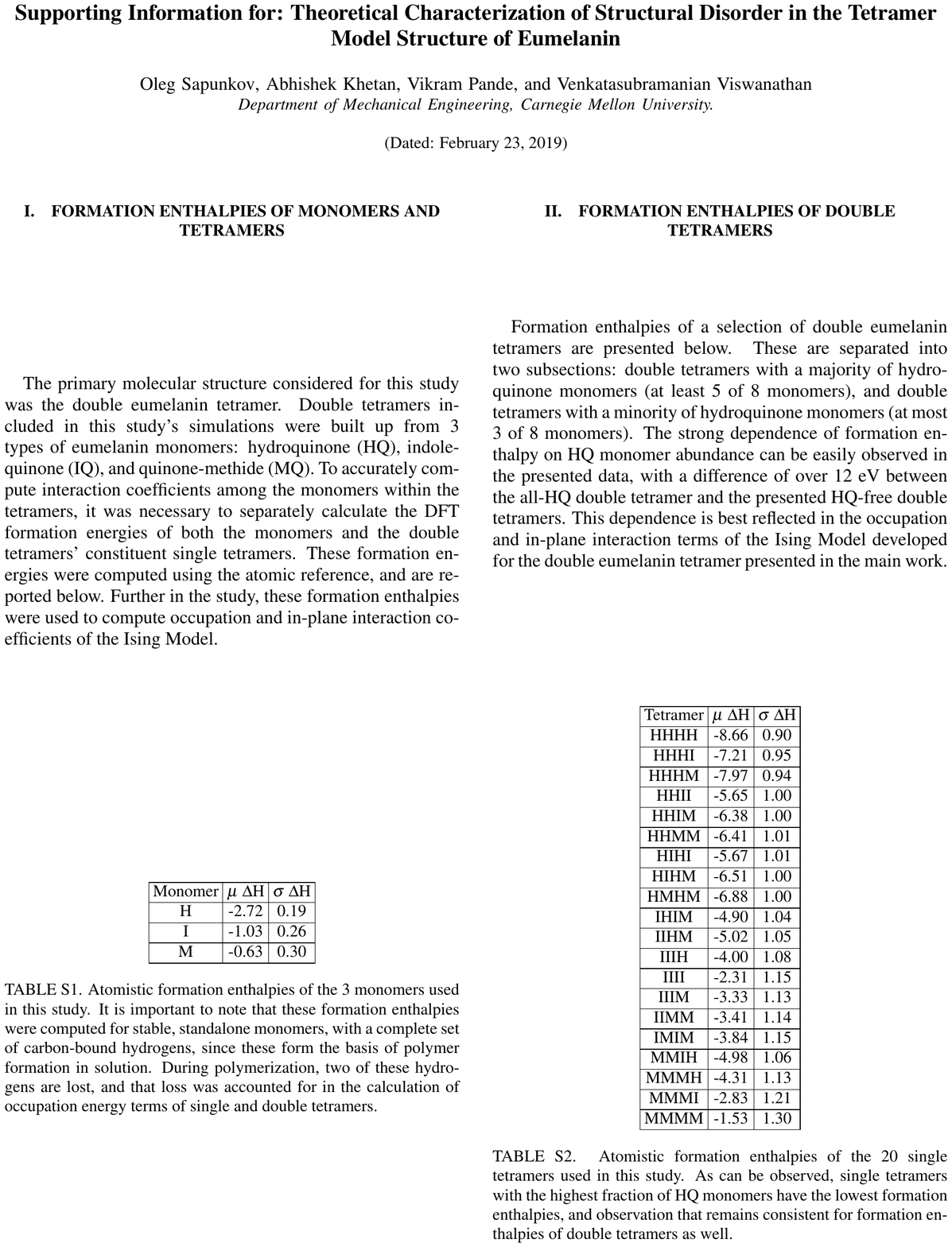}
}

\end{document}